\documentclass[11pt]{article}
\usepackage{amssymb,amsfonts,amsmath,amsthm}
\usepackage{amsgen}
\usepackage{amscd}
\usepackage{graphicx}
\usepackage[algo2e]{algorithm2e}
\usepackage{algorithm}
\usepackage{multirow}
\usepackage{extramarks}
\usepackage{fancyhdr}
\usepackage{subfigure}
\usepackage[bookmarks=false]{hyperref}
\usepackage{tikz}
\usepackage{epstopdf}
\input{xy}
\xyoption{all}
\theoremstyle{plain}

\theoremstyle{remark}

\numberwithin{equation}{section}
\numberwithin{algorithm}{section}

\begin{document}

\title{Modeling Stochasticity and Variability in Gene Regulatory Networks}

\author{David Murrugarra$^{a,b}$, Alan Veliz-Cuba$^{c}$, Boris Aguilar$^{d}$, Seda Arat$^{a,b}$, \\
Reinhard Laubenbacher$^{a,b}$}
\maketitle
{\footnotesize
     \centerline{$^a$Department of Mathematics,
      Virginia Polytechnic Institute and State University,}
  \centerline{Blacksburg, VA 24061-0123, USA}
}
{\footnotesize
    \centerline{$^b$Virginia Bioinformatics Institute,
      Virginia Polytechnic Institute and State University,}
  \centerline{Blacksburg, VA 24061-0477, USA}
}
{\footnotesize
  \centerline{$^c$Department of Mathematics,
  University of Nebraska Lincoln,}
  \centerline{Lincoln, NE 68588, USA}
}
{\footnotesize
  \centerline{$^d$Department of Computer Science,
  Virginia Polytechnic Institute and State University,}
  \centerline{Blacksburg, VA 24061-0123, USA}
}
\begin{center}
Abstract
\end{center}
Modeling stochasticity in gene regulatory networks is an important and complex problem in molecular systems biology. To elucidate intrinsic noise, several modeling strategies such as the Gillespie algorithm have been used successfully. This paper contributes an approach as an alternative to these classical settings. Within the discrete paradigm, where genes, proteins, and other molecular components of gene regulatory networks are modeled as discrete variables and are assigned a logical rules describing their regulation through interactions with other components. Stochasticity is modeled at the biological function level under the assumption that even if the expression levels of the input nodes of an update rule guarantee activation or degradation there is a probability that the process will not occur due to stochastic effects. This approach allows a finer analysis of discrete models and provides a natural setup for cell population simulations to study cell-to-cell variability. We applied our methods to two of the most studied regulatory networks, the outcome of lambda phage infection of bacteria and the p53-mdm2 complex.
\section{Introduction}

Variability at the molecular level, defined as the phenotypic differences within a genetically identical population of cells exposed to the same environmental conditions, has been observed experimentally~\cite{Elowitz,Acar,St-Pierre,Geva-Zatorsky}. Understanding mechanisms that drive variability in molecular networks is an important goal of molecular systems biology, for which mathematical modeling can be very helpful. 
Different modeling strategies have been used for this purpose and, 
depending on the level of abstraction of the mathematical models, there are several
ways to introduce stochasticity. Dynamic mathematical models can be broadly divided into two classes: continuous, such as systems of differential equations (and their stochastic variants) and discrete, such as Boolean networks and their generalizations (and their stochastic variants). This paper will focus on stochasticity and discrete models.

Discrete models do not require detailed information about kinetic rate constants and they tend to be more intuitive. In turn, they only provide qualitative information about the system. The most general setting is as follows. Network nodes represent genes, proteins, and other molecular components of gene regulation, while network edges describe biological interactions among network nodes that are given as logical rules representing their interactions. Time in this framework is implicit and progresses in
discrete steps. More formally, let $x_1,\ldots , x_n$ be variables, which can take values in finite sets $X_1,\ldots , X_n$, respectively.  Let $X = X_1\times\cdots\times X_n$ be the Cartesian product. 
A discrete dynamical system in the variables $x_1, \ldots , x_n$ is a function 
\begin{displaymath}
f = (f_1,\dots,f_n):X\rightarrow X
\end{displaymath}   
where each coordinate function $f_i: X\rightarrow X_i$ is a function in a subset of $\{x_1,\dots,x_n\}$. 
Dynamics is generated by iteration of $f$, and different update schemes can be used for this purpose. 
As an example, if $X_i = \{0, 1\}$ for all $i$, then each $f_i$ is a Boolean rule and $f$ is a Boolean network where all the variables are updated simultaneously. We will assume that each $X_i$ comes with a natural total ordering
of its elements (corresponding to the concentration levels of the associated molecular species). 
Examples of this type of dynamical system representation are Boolean networks, logical models and Petri nets~\cite{ThomasD'Ari,Chaouiya,Irons}. 

To account for stochasticity in this setting several methods have been considered. Probabilistic Boolean networks~\cite{Shmulevich2002,Shmulevich2010} introduce stochasticity in the update functions, allowing a different update function to be used at each iteration, chosen from a probability space of such functions for each
network node. Other approaches include~\cite{Shunsuke,Garg,Ribeiro2007}. 
These models will be discussed in more detail in the next section. In this paper we present a model type related
to probabilistic Boolean networks, with additional features. We show that this model type is 
natural and a useful way to simulate gene regulation as a stochastic process, and is very useful to simulate experiments with cell populations. 
 
\subsection{Modeling Stochasticity in Gene Regulatory Networks}
Gene regulation processes are inherently stochastic. Accurately modeling this stochasticity is a complex and important goal in molecular system biology. Depending on the level of knowledge of the biological system and the availability of data for it one could follow different approaches. 
For instance, viewing a gene regulatory network as a biochemical reaction network,
the Gillespie algorithm can be applied to simulate each biochemical reaction separately generating a random walk corresponding to a solution of the chemical master equation of the system~\cite{Gillespie,Gillespie2007}. 
At an even more detailed level one could introduce time delays into the Gillespie simulations to 
account for realistic time delays in activation or degradation such as in 
circadian rhythms~\cite{Bratsun, Ribeiro2010, Ribeiro2006}. 
At a higher level of abstraction, stochastic differential equations~\cite{Toulouse} contain a 
deterministic approximation of the system and an additional random white noise term. 
However, all these schemes require that all the kinetic rate constants to be known 
which could represent a strong constraint due to the difficulty of measuring kinetic parameters, limiting 
these approaches to small systems.

As mentioned in the introduction, discrete models are an alternative to continuous models, 
which do not depend on rate constants. In this setting, several approaches to 
introduce stochasticity have been proposed. Specially for Boolean networks, stochasticity has been introduced by flipping node states from 0 to 1 or vice versa with some flip probability~\cite{Ribeiro2007,Alvarez,Davidich,Willadsen}. However, it has been argued that this way of introducing stochasticity into the system usually leads to over-representation of noise~\cite{Garg}. The main criticism of this approach is that it does not take into consideration the correlation between the expression values of input nodes and the probability of flipping the expression of a node due to noise. In fact, this approach models the stochasticity at a node regardless of the susceptibility to noise of the underlying biological function~\cite{Garg}. 

Probabilistic Boolean networks (PBNs)~\cite{Shmulevich2002,Shmulevich2010,Dougherty} is another stochastic method proposed within the discrete strategy. PBNs model the choice among alternate biological functions during the iteration process, rather than modeling the stochasticity of the function failure itself. We have adopted a special case of this
setting, in which every node has associated to it two functions: the function that governs its evolution over
time and the identity function. If the first is chosen, then the node is updated based on its logical rule. 
When the identity function is chosen, then the state of the node is not updated. The key difference to 
a PBN is the assignment of probabilities that govern which update is chosen. In our setting, each function gets assigned
two probabilities. Precisely, let $x_i$ be a variable. We assign to it a probability $p_i^\uparrow$, which 
determines the likelihood that $x_i$ will be updated based on its logical rule, if this update leads to an increase/activation of the variable. Likewise, a probability $p_i^\downarrow$ determines this probability in case the variable is decreased/inhibited. The necessity for considering two different probabilities is that activation and degradation represent different biochemical processes and even if these two are encoded by the same function, their propensities in general are different. This is very similar to what is considered in differential equations modeling, where,
for instance, the kinetic rate parameters for activation and for degradation/decay are, in principle, different.

Note that all these approaches only take account of intrinsic noise which is generated from small fluctuations in concentration levels, small number of reactant molecules, and fast and slow reactions. Another source of stochasticity is related to extrinsic noise such as a noisy cellular environment and temperature. For more about intrinsic vs extrinsic noise see~\cite{Swain-Elowitz,St-Pierre}.

\section{Method}

Our aim is to model stochasticity at the biological function level under the main assumption that even if the expression levels of the input nodes of an update function guarantee activation or degradation there is a probability that the process will not occur due to stochasticity, for instance, if some of the chemical reactions encoded by the update function may fail to occur. This is similar to models based on the chemical master equation. This model type introduces activation and degradation propensities. More formally, let $x_1, \ldots , x_n$ be variables which can take values in finite sets $X_1,\ldots , X_n$, respectively. Let $X = X_1\times\cdots\times X_n$ be the Cartesian product. Thus, the formal definition of a 
\emph{stochastic discrete dynamical system} in the variables $x_1, \ldots , x_n$ is a collection of $n$ triplets
\begin{displaymath}
F=\{f_i,p_i^\uparrow,p_i^\downarrow\}^n_{i=1}
\end{displaymath}
where  
\begin{itemize}
  \item $f_i : X\rightarrow X_i$ is the update function for $x_i$, for all $i = 1,\dots,n$.
  \item $p_i^\uparrow$ is the activation propensity.
  \item $p_i^\downarrow$ is the degradation propensity.
  \item $p_i^\uparrow,p_i^\downarrow\in[0,1]$. 
\end{itemize}
We now proceed to study the dynamics of such systems and two specific models as illustration. 
   
\subsection{Dynamics of Stochastic Discrete Dynamical Systems}
Let $F=\{f_i,p_i^\uparrow,p_i^\downarrow\}^n_{i=1}$ be a stochastic discrete dynamical system and consider $x\in X$. For all $i$ we define  $\pi_{i,x}(x_i\rightarrow f_i(x))$  and $\pi_{i,x}(x_i\rightarrow x_i)$ by
\[
\pi_{i,x}(x_i\rightarrow f_i(x))=
\begin{cases}
p_i^\uparrow, & \text{if $x_i<f_i(x)$},\\
p_i^\downarrow, & \text{if $x_i>f_i(x)$},\\
1, & \text{if $x_i=f_i(x)$}.
\end{cases}
\]
\[
\pi_{i,x}(x_i\rightarrow x_i)=
\begin{cases}
1-p_i^\uparrow, & \text{if $x_i<f_i(x)$},\\
1-p_i^\downarrow, & \text{if $x_i>f_i(x)$},\\
1, & \text{if $x_i=f_i(x)$}.
\end{cases}
\]
That is, if the possible future value of the $i$-th coordinate is larger (smaller, resp.) than the current value, then the activation (degradation) propensity determines the probability that the $i$-th coordinate will increase (decrease) its current value. If the $i$-th coordinate and its possible future value are the same, then the $i$-th coordinate of the system will maintain its current value with probability 1. Notice that $\pi_{i,x}(x_i\rightarrow y_i)=0$ for all $y_i\notin \{x_i,f_i(x)\}$.

The dynamics of $F$ is given by the weighted graph $X$ which has an edge from $x\in X$ to $y\in X$ if and only if $y_i\in\{x_i,f_i(x)\}$ for all $i$. The weight of an edge $x\rightarrow y$ is equal to the product 
\begin{displaymath}
w_{x\rightarrow y}=\prod_{i=1}^n \pi_{i,x}(x_i\rightarrow y_i)
\end{displaymath} 

By convention we omit edges with weight zero. See the Supporting Materials for 
pseudocodes of algorithms to compute dynamics of stochastic discrete dynamical systems. Software to test examples is available at \url{http://dvd.vbi.vt.edu/adam.html} \cite{ADAM}
as a web tool (choose Discrete Dynamical Systems (SDDS) in the model type).

Given $F=\{f_i,p_i^\uparrow,p^\downarrow_i\}^n_{i=1}$ a stochastic discrete dynamical system, it is straightforward to verify that $F$ has the same steady states (fixed points) as the deterministic system $G=\{f_i\}^n_{i=1}$ (see supporting material). It is also important to note that the dynamics of $F$ includes the different trajectories that can be generated from $G$ using other common update mechanisms such as the synchronous and asynchronous schemes (see supporting material). 
\subsubsection{Example}

Let $n=2$, $X = \{0,1\}\times\{0,1\}$, $F=(f_1,f_2):X\rightarrow X$, where
\begin{center}
\begin{tabular}{| c | c || c | c |}
\hline
$x_1$ & $x_2$ & $f_1$ & $f_2$\\ \hline
   0& 0 & 0 & 0\\
   0& 1 & 1 & 0\\ 
   1& 0 & 0 & 1\\ 
   1 & 1& 1 & 0\\
\hline
\end{tabular}
\end{center}
 
and

\begin{center}
\begin{tabular}{| c  c  c |}
\hline
   & $x_1$ & $x_2$\\ \hline
Activation& .1 & .5 \\ 
\hline
Degradation&.2& .9 \\ \hline
\end{tabular}
\end{center}

\tikzstyle{place}=[circle,draw=blue!50,fill=blue!20,thick,
inner sep=0pt,minimum size=6mm]
\tikzstyle{transition}=[rectangle,draw=black!50,fill=black!20,thick,
inner sep=0pt,minimum size=4mm]

\begin{center}
\begin{tikzpicture}[->,>=stealth,shorten >=1pt,auto,node distance=2.8cm,semithick]
\node[place] (A) {01};
\node[place] (B) [below of=A] {10};
\node[place] (C) [right of=B] {00};
\node[place] (D) [left of=B] {11};
\path (A) edge [loop above] node {$9\%$} (A)
          edge  [bend right] node [swap] {$9\%$} (B)
          edge [bend left] node {$81\%$} (C)
          edge [bend right] node [swap] {$1\%$} (D);
\path (B) edge [loop below] node {$40\%$} (B)
          edge [bend right] node [swap] {$10\%$} (A)
          edge node {$10\%$} (C)
          edge [bend right] node {$40\%$} (D);          
\path (C) edge [loop below] node {$100\%$} (C);
\path (D) edge [loop below] node {$10\%$} (D)
          edge  [bend right] node [swap] {$90\%$} (B);
\end{tikzpicture}
\end{center}
\begin{displaymath}
Pr(01\rightarrow10) = (.1)(.9)=.09,\ Pr(01\rightarrow00) = (1-.1)(.9) = .81
\end{displaymath}
\begin{displaymath}
Pr(01\rightarrow01) = (1-.1)(1-.9)=.09,\ Pr(01\rightarrow11) = (.1)(1-.9)=.01
\end{displaymath}
\begin{displaymath}
Pr(10\rightarrow10) = (1 - .2)(1 - .5)=.4,\ Pr(10\rightarrow01) = (.2)(.5) = .1
\end{displaymath}
\begin{displaymath}
Pr(10\rightarrow00) = (.2)(1-.5)=.1,\ Pr(10\rightarrow11) = (1-.2)(.5)=.4
\end{displaymath}
\begin{displaymath}
Pr(11\rightarrow11) = (1)(1-.9)=.1,\ Pr(11\rightarrow10) = (1)(.9)=.9
\end{displaymath}
\begin{displaymath}
Pr(00\rightarrow00) = (1)(1) = 1.
\end{displaymath}
This figure shows that there is a 9\% chance that the system will transition from $01$ to $10$. Similarly, there is an 81\% chance that the system will transition from $01$ to $00$. The latter was expected because there is a high degradation propensity for $f_2$. Note that $00$ is a fixed point, i.e., there is $100\%$ chance of staying at this state. 
\section{Applications}
We illustrate the advantages of this model type by applying it to two widely studied biological systems,
the regulation of the p53-mdm2 network and the control of the outcome of phage lambda infection of bacteria. These regulatory networks were selected because stochasticity plays a key role in their dynamics.
\subsection{Regulation in the p53-mdm2 network}
\begin{figure}
\begin{center}
\includegraphics[width = \textwidth]{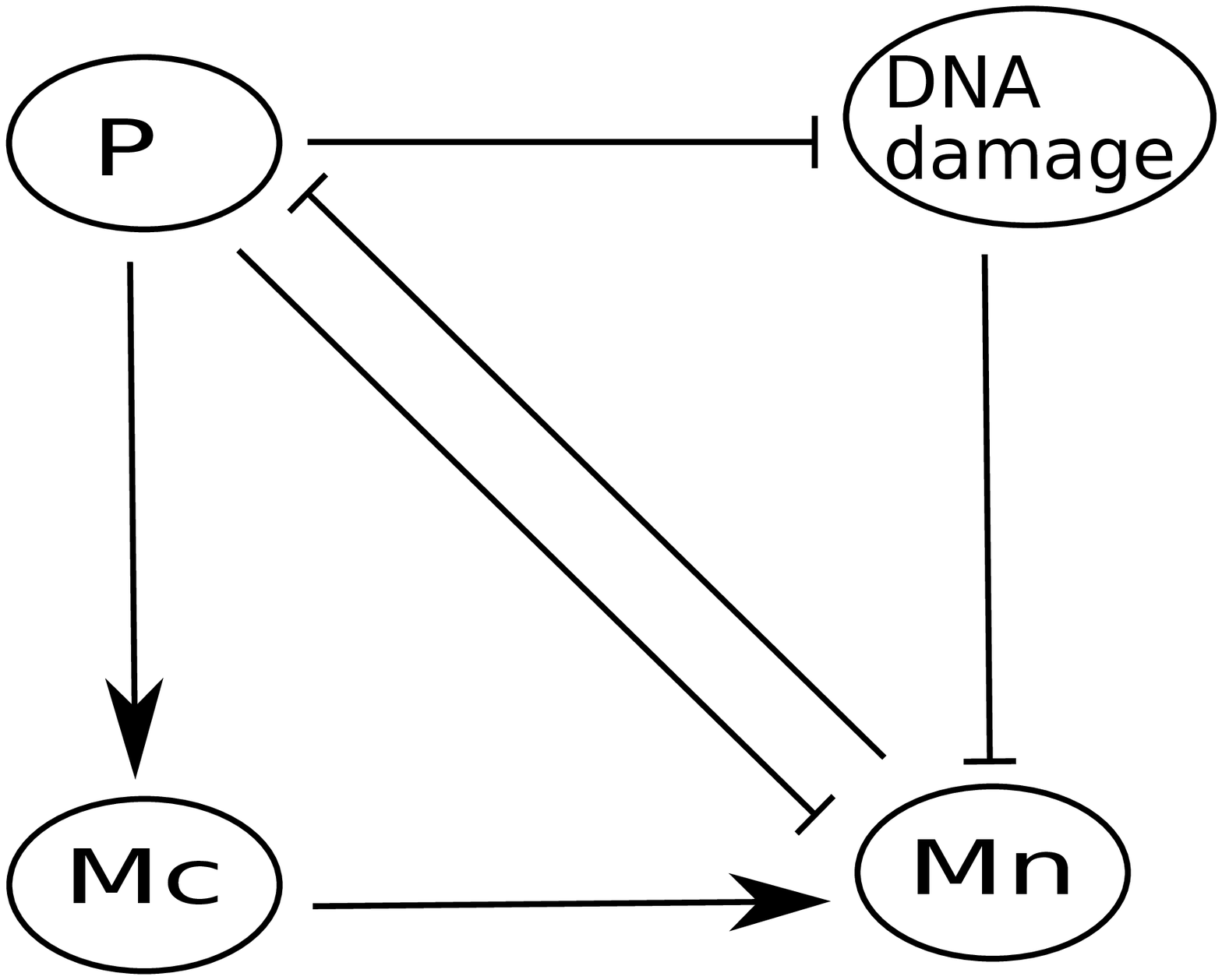}
\caption{Four-variable model for the p53-Mdm2 regulatory network. P, Mc, and Mn stand for protein p53, cytoplasmic Mdm2, and nuclear Mdm2 respectively.}
\label{p53WiringDiagram}
\end{center}
\end{figure}

The p53-Mdm2 network is one of the most widely studied gene regulatory networks.  W.~Abou-Jaude, D.~Ouattara, and M.~Kauffman~\cite{Abou-Jaoude} proposed a logical four-variable model to describe the dynamics of the tumor suppressor protein p53 and its negative regulator Mdm2 when DNA damage occurs. The wiring diagram of this model is represented in Figure~\ref{p53WiringDiagram}, where P denotes cytoplasmic p53, nucleic p53, and the gene p53. Mc and Mn stand for cytoplasmic Mdm2 and nuclear Mdm2, respectively. DNA damage caused by ionic irradiation decreases the level of nucleic Mdm2 which enables p53 to accumulate and to remain active, playing a key role in reducing the effect of the damage. There is a negative feedback loop involving three components: p53 increases the level of cytoplasmic Mdm2 which, in turn, increases the level of nuclear Mdm2. Nucleic Mdm2 reduces p53 activity. This model also contains a positive feedback loop involving two components where p53 inhibits its negative regulator nucleic Mdm2. Note the dual role of P, as it positively regulates nucleic Mdm2 through cytoplasmic Mdm2. On the other hand, P negatively regulates nucleic Mdm2 by inhibiting Mdm2 nuclear translocation~\cite{Abou-Jaoude}. For more about the p53-Mdm2 system, see~\cite{Abou-Jaoude,Batchelor, Geva-Zatorsky}.

\begin{figure}
\begin{center}
\includegraphics[width=\textwidth]{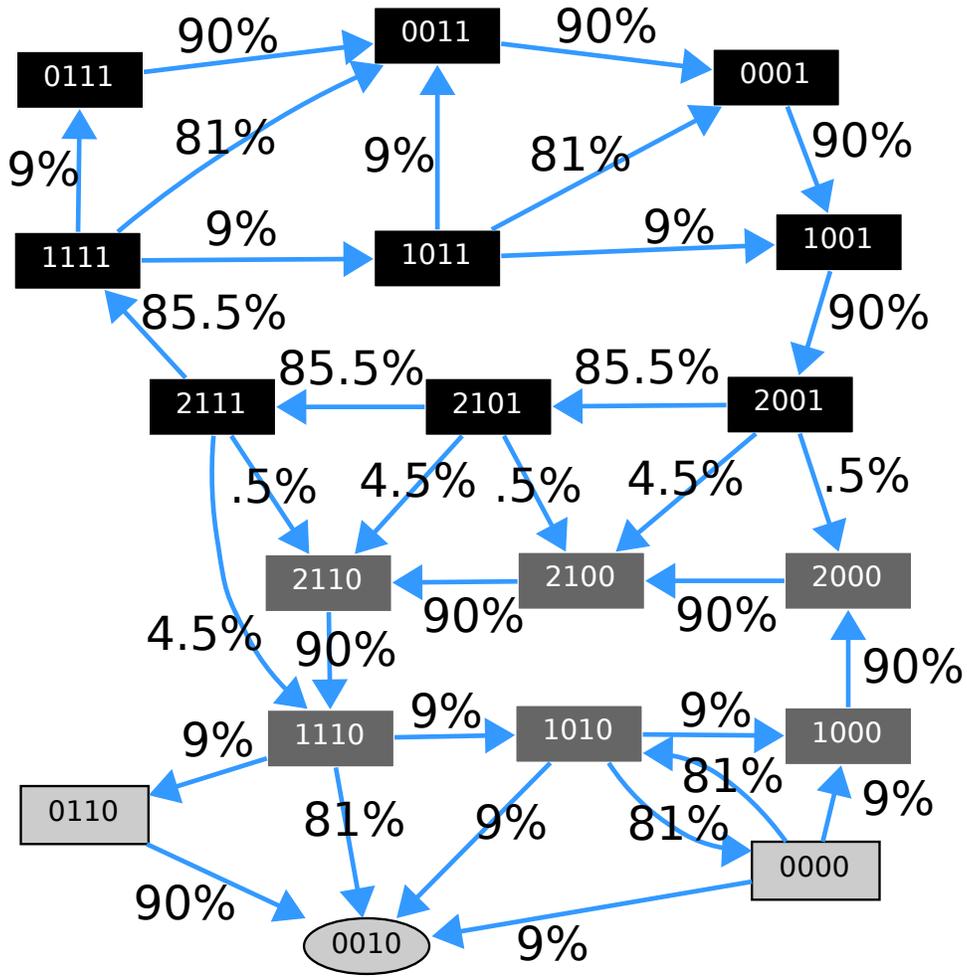}
\caption{State space diagram for parameters described in Table~\ref{delay_matrix_p53}. The numbers next to the edges encode the transition probabilities. The order of the variables in each vector state is P, Mc, Mn, DNA damage. Self-loops are not depicted. States with darker background comprise the cycle with DNA damage. A second cycle with a lighter shaded background corresponds to the cycle with no DNA damage. The oval shaped state is a steady state.}
\label{p53_state_space}
\end{center}
\end{figure}

The dynamic behavior of the system is represented in a network of transitions called its state space, see Figure~\ref{p53_state_space}. This specifies the different paths to follow and the probabilities of following a specific trajectory from a given state. Dynamics here is not deterministic, i.e., most of the state vectors have different trajectories they
can follow. The propensity parameters in Table~\ref{delay_matrix_p53} determine the likelihood of following certain paths. The state 0010 is a steady state, which is differentiated from the others by its oval shape.

The state space for this model is specified by $[0,2]\times[0,1]\times[0,1]\times[0,1]$, that is, except for the first variable $P$ which has three levels $\{0,1,2\}$, all the other variables are Boolean. The update functions for this model are provided in the supporting material and also in the model repository of our web tool at \url{http://dvd.vbi.vt.edu/adam.html}.

\begin{figure}
\begin{center}
\includegraphics[width=\textwidth]{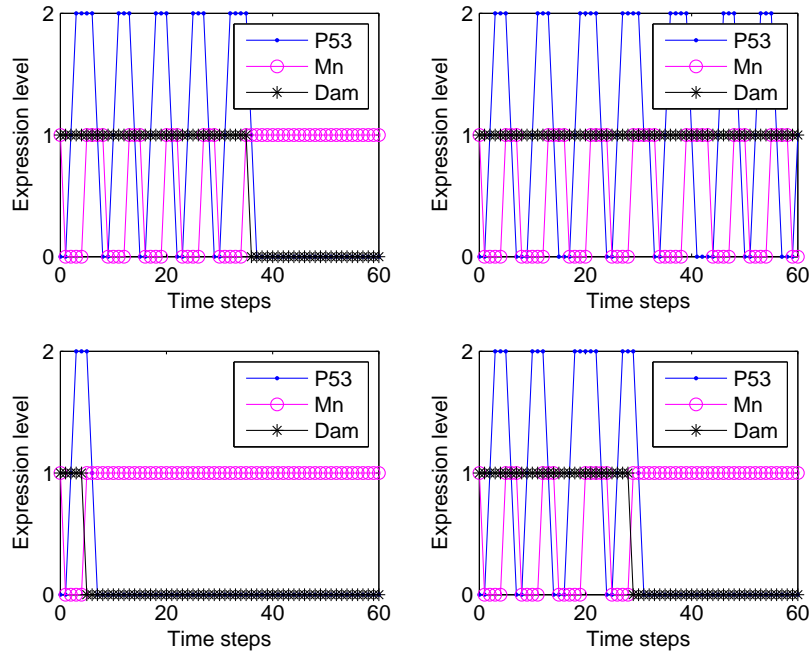}
\caption{Individual cell simulations for parameters described in Table~\ref{delay_matrix_p53}. Each subfigure shows oscillations as long as the damage is present. This figure shows variability in the timing of damage repair and in the period of the oscilations. Each frame was generated from a single simulation with sixty time steps. The x-axis represents discrete time steps and the y-axis the expression level. The initial state for all simulations is $0011$.}
\label{single_cell_p53}
\end{center}
\end{figure}

\begin{figure}
\begin{center}
\includegraphics[width=\textwidth]{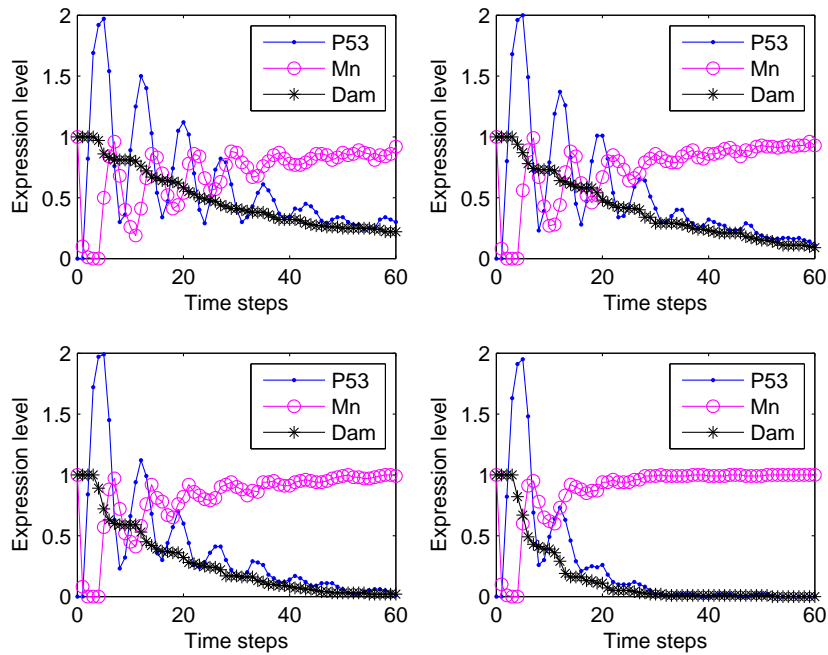}
\caption{Cell population simulations. Each subfigure was generated from one hundred simulations, each representing a single cell with sixty time steps. Starting from the top left frame to the right bottom frame the degradation propensity for DNA damage was increased by 5\%, i.e. $p_{dam}^\downarrow=.05$ (top left), $p_{dam}^\downarrow=.10$ (top right), $p_{dam}^\downarrow=.15$ (bottom left), and $p_{dam}^\downarrow=.2$ (bottom right). The x-axis represents discrete time steps and the y-axis the average expression level. The initial state for all simulations was $0011$. This figure shows that, if the intensity of the damage is increased more cells exhibit oscillations in the level of p53, in agreement with experimental observations~\cite{Geva-Zatorsky}.}
\label{population_cells_p53}
\end{center}
\end{figure}

Individual cell simulations render plots similar to the ones shown in Figure~\ref{single_cell_p53}. Each subfigure shows oscillations as long as the damage is present with a variability in the timing of damage repair. On the other hand, cell population simulations, Figure~\ref{population_cells_p53}, exhibit damped oscillations of the expression level of p53 as the degradation propensities of the damage increases. This is correlated with the fact that, if the intensity of the damage is increased, more cells exhibit oscillations in the level of p53 which was experimentally observed in~\cite{Geva-Zatorsky}. The initial state for all simulations was $0011$ which represents the state when DNA damage is introduced ($0010$ is the steady state without perturbation).

To highlight the features of our approach we compare our model with the one presented in~\cite{Abou-Jaoude} in which variability has been analyzed. The main difference between these two models is the way the simulations are performed. In~\cite{Abou-Jaoude}, the transition from one state to the next is determined by parameters called ``on" and ``off" time delays. For instance, to transition from $2001$ to $2101$ it is required that $t_{Mc}<t_{\overline{dam}}$ which means that the ``on" delay for $Mc$ (time for activating) is less than the ``off" delay (time for degrading) of the damage. Otherwise, if $t_{Mc} > t_{\overline{dam}}$ the system will transition from $2001$ to $2000$. In this paper, transitions from one state to others are given as probabilities which are determined from the propensity probabilities. Therefore, the complexity of the model presented here is at the level of the wiring diagram (i.e. the number of variables) while the complexity of the model in~\cite{Abou-Jaoude} is at the level of the state space (i.e. number of possible states) which is exponential in the number of variables. Another key difference is the way DNA damage repair is modeled. In~\cite{Abou-Jaoude}, a delay parameter $t_{\overline{dam}}$ is associated with the disappearance of the damage, and this is decreased by a certain amount $\tau$ at each iteration so that $t_{\overline{dam}}^{(n)} = t_{\overline{dam}}^{(0)} - n\tau\geqslant0$ where $n$ is the number of iterations. In order to simulate DNA damage with this approach it is required to estimate $\tau, n$, and $t_{\overline{dam}}^{(0)}$. Within our model framework a single parameter, the degradation propensity, is used to model the damage repair which is a more natural setup. 

\begin{table}
  \centering
\begin{tabular}{| c | c | c | c | c |}
\hline
   & $P$ & $Mc$ & $Mn$ & $Dam$\\ \hline
Activation   & .9 & .9 & .9 & 1 \\ \hline
Degradation   & .9 & .9 & .9 & .05 \\
\hline
\end{tabular}
  \caption{Propensity probabilities for the p53-mdm2 regulatory network. Note that there is a low degradation propensity for DNA damage.}\label{delay_matrix_p53}
\end{table}

\subsection{Phage lambda infection of bacteria}

\begin{figure}
\begin{center}
\includegraphics[width=\textwidth]{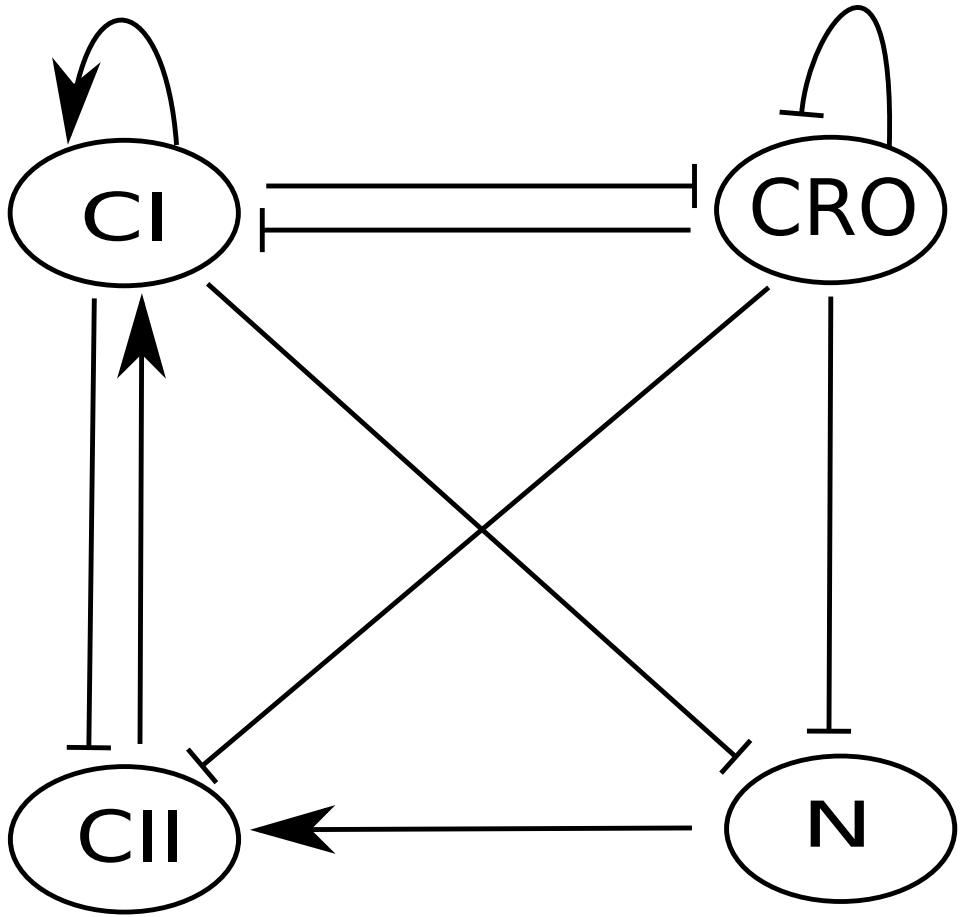}
\caption{Wiring diagram for phage lambda infection model.}
\label{Phage4White}
\end{center}
\end{figure}

Control of the outcome of phage lambda infection is one of the best understood regulatory systems~\cite{Ptashne,ThieffryThomas, St-Pierre}.  Figure~\ref{Phage4White} depicts its core regulatory network that was first modeled by Thieffry and Thomas~\cite{ThieffryThomas} using a logical approach. This model encompasses the roles of the regulatory genes \textit{CI}, \textit{CRO}, \textit{CII}, and \textit{N}. From experimental reports~\cite{ThieffryThomas, Reichardt, Kourilsky, St-Pierre} it is known that, if the gene CI is fully expressed, all other genes are off. In the absence of CRO protein, CI is fully expressed (even in the absence of N and CII). CI is fully repressed provided that CRO is active and CII is absent. 

\begin{figure}
\begin{center}
\includegraphics[width=\textwidth]{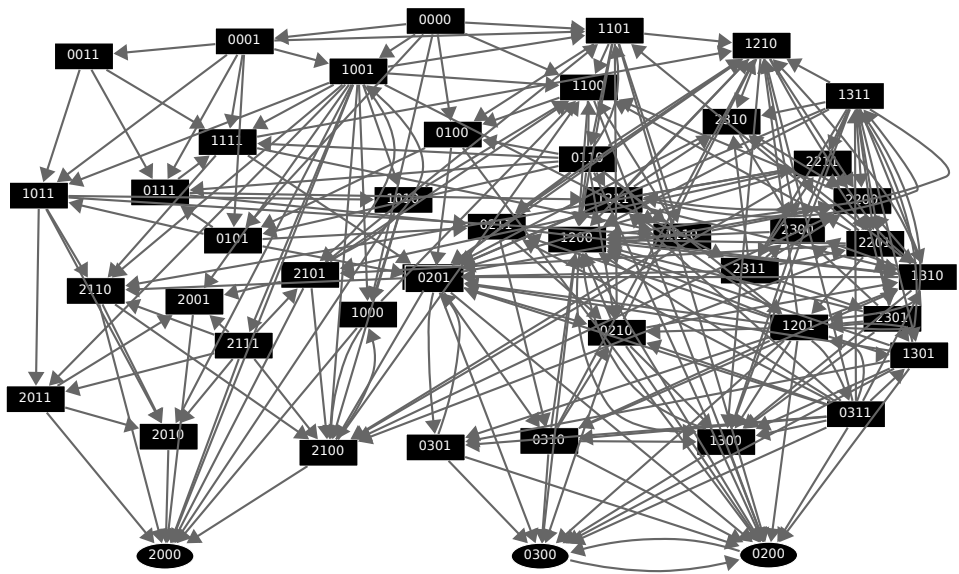}
\caption{State space for phage lambda model. The order of variables in each vector state is $CI, CRO, CII, N$. The steady state $2000$ represents lysogeny where $CI$ is fully expressed while other genes are off. The cycle between $0200$ and $0300$ represents lysis where $CRO$ is active and other genes are repressed. Self-loops are not depicted.}
\label{Lambda_state_space}
\end{center}
\end{figure}

The dynamics of this network is a bistable switch between lysis and lysogeny, Figure~\ref{Lambda_state_space}. Lysis is the state where the phage will be replicated, killing the host. Otherwise, the network will transition to a state called lysogeny where the phage will incorporate its DNA into the bacterium and become dormant. It has been suggested~\cite{Arkin, ThieffryThomas} that these cell fate differences are due to spontaneous changes in the timing of individual biochemical reaction events. 

The state space for this model is specified by $[0,2]\times[0,3]\times[0,1]\times[0,1]$, that is, the first variable, $CI$, has three levels $\{0,1,2\}$, the second variable, $CRO$, has four levels $\{0,1,2,3\}$, and the third and fourth variables, $CII$ and $N$, are Boolean. Update functions for this model are available in our supporting material. This model has a steady state, $2000$, and a 2-cycle involving $0200$ and $0300$. The steady state $2000$ represents lysogeny where $CI$ is fully expressed while the other genes are off. The cycle between $0200$ and $0300$ represents lysis where $CRO$ is active and other genes are repressed. 

\begin{figure}
\begin{center}
\includegraphics[width=\textwidth]{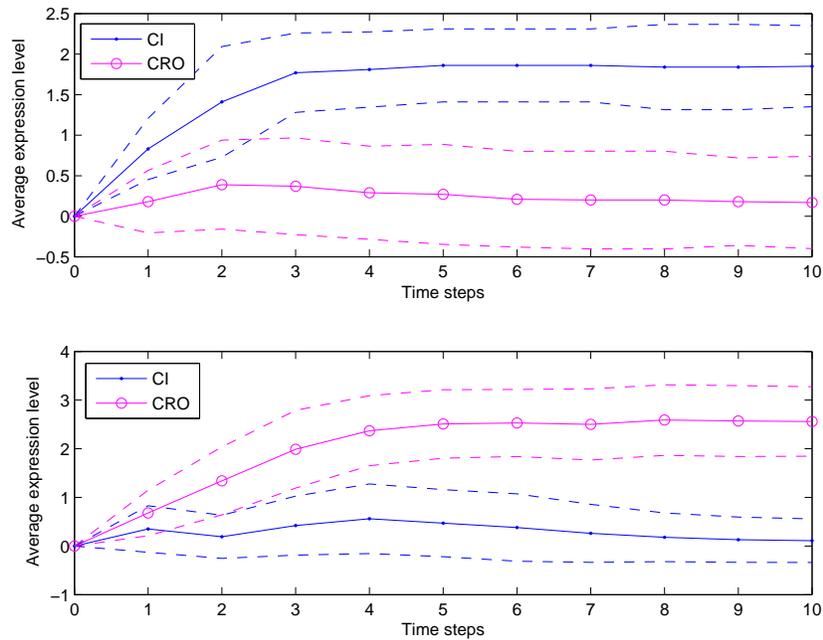}
\caption{Cell population simulations. Both figures were generated from one hundred simulations, each representing a single cell iteration of ten time steps. Top frame for parameters in Table~\ref{delay_matrix_lysogeny} shows 93\% lysis and 7\% lysogeny while bottom frame for parameters in Table~\ref{delay_matrix_lysis} shows 4\% lysis and 96\% lysogeny. The x-axis represents discrete time steps while the y-axis shows the average expression level. The initial state for all the simulations is $0000$. Solid (circle) points correspond to the average of $CI$ ($CRO$), and dashed lines represent standard deviations.}
\label{lysis_lysogeny}
\end{center}
\end{figure}

\begin{figure}
\begin{center}
\includegraphics[width=\textwidth]{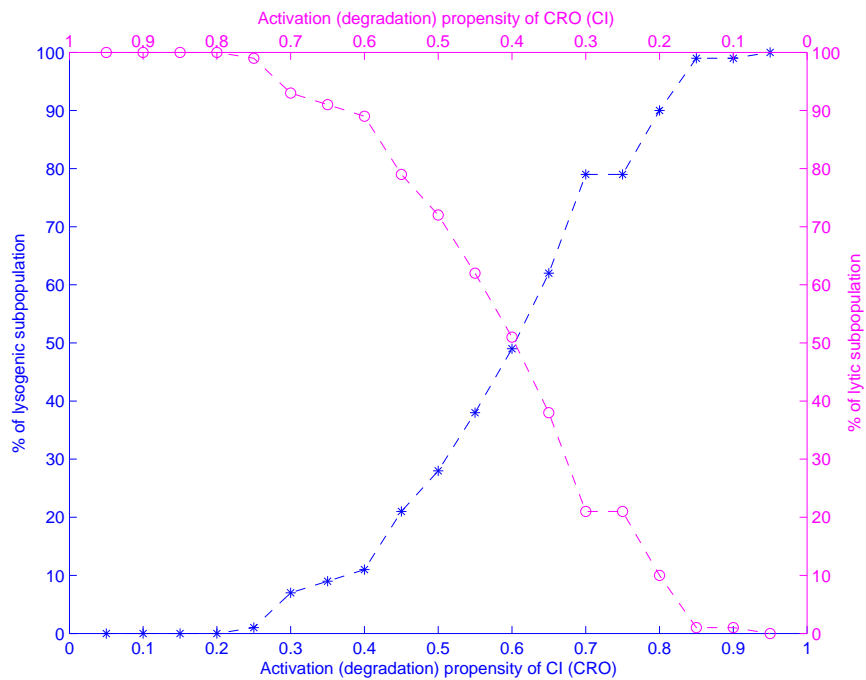}
\caption{Variation in developmental outcome as a function of the propensity parameters. Star points indicate the percentage of networks that transition to lysogeny and circle shaped points indicate the percentage of networks that end up in lysis. Bottom axis represents the activation (and degradation) propensities for $CI$ ($CRO$) in increasing order. Likewise, the top axis represents the activation (and degradation) propensities for $CRO$ ($CI$) in decreasing order.}
\label{lambda4_double_axes}
\end{center}
\end{figure}

Cell population simulations were performed to measure the cell-to-cell variability. Figure~\ref{lysis_lysogeny} was generated using the probabilities given in Table~\ref{delay_matrix_lysogeny} (top frame) and Table~\ref{delay_matrix_lysis} (bottom frame). The x-axis in both subfigures represents discrete time steps while the y-axis 
captures the average expression level. The initial state for all simulations was $0000$ which represents the state of the bacterium at the moment of phage infection. Figure~\ref{lysis_lysogeny} shows variability in developmental outcome, some of the networks transition to lysis while others transition to lysogeny. 
To measure how sensitive the dynamics of the network is to changes in the propensity probabilities we have plotted the outcome of lysis-lysogeny percentages for different choices of these parameters. Figure~\ref{lambda4_double_axes} shows the variation in developmental outcome as a function of the propensity parameters of $CI$ and $CRO$. Star points indicate the percentage of networks that transition to lysogeny and circle shaped points indicate the percentage of networks that end up in lysis. The bottom x-axis contains activation propensities for $CI$ and degradation propensities for $CRO$ while the top x-axis contains activation propensities for $CRO$ and degradation propensities for $CI$. The activation and degradation propensities for $CII$ and $N$ were all set equal to $.9$. Although the probability distributions for $CI$ and $CRO$ are very symmetric in Figure~\ref{lambda4_double_axes}, it gives a good idea of how the variability in developmental outcome will change as the propensity parameters change. 
\begin{table}
  \centering
\begin{tabular}{| c | c | c | c | c |}
\hline
   & $CI$ & $CRO$ & $CII$ & $N$\\ \hline
Activation       & .8 & .2 & .9 & .9 \\ \hline
Degradation   & .2 & .8 & .9 & .9 \\
\hline
\end{tabular} 
  \caption{Propensity parameters  for Figure~\ref{lysis_lysogeny} (top frame). There is a high activation propensity for $CI$ while a low activation propensity for $CRO$.}\label{delay_matrix_lysogeny}
\end{table}

\begin{table}
  \centering
\begin{tabular}{| c | c | c | c | c |}
\hline
   & $CI$ & $CRO$ & $CII$ & $N$\\ \hline
Activation       & .3 & .7 & .9 & .9 \\ \hline
Degradation   & .7 & .3 & .9 & .9 \\
\hline
\end{tabular} 
  \caption{Propensity parameters for Figure~\ref{lysis_lysogeny} (bottom frame). There is a high activation propensity for $CRO$ while a low activation propensity for $CI$.}\label{delay_matrix_lysis}
\end{table}
\section{Discussion}

Using a discrete modeling strategy, this paper introduces a framework to simulate stochasticity in gene regulatory networks at the function level, based on the general concept of probabilistic Boolean networks. 
It accounts for intrinsic noise due to spontaneous differences in timing, small fluctuations in concentration levels, small numbers of reactant molecules, and fast and slow reactions. This framework was tested using two widely studied regulatory networks, the regulation of the $p53-mdm2$ network and the control of phage lambda infection of bacteria. It is shown that in both of these examples the use of propensity probabilities 
for activation and inhibition of network nodes provides a natural setup for cell population simulations to study cell-to-cell variability. The new features of this framework are the introduction of activation and degradation propensities that determine how fast or slow the discrete variables are being updated. This provides the ability to generate more realistic simulations of
both single cell and cell population dynamics. In the example of the $p53-mdm2$ system, one can see that individual simulations show sustained oscillations when DNA damage is present, while at the cell population level
these individual oscillations average to a damped oscillation. This agrees with experimental observations~\cite{Geva-Zatorsky}.  In the second example, $\lambda$-phage infection of bacteria, it is observed that differences in developmental outcome due to intrinsic noise can be captured with this framework. 
Due to the lack of experimental data we are unable to calibrate the model so that it reproduces the correct difference in percentages due to intrinsic noise. So instead we present a plot of the difference in developmental outcome as a function of the propensity parameters. 

It is worth noting that this paper addresses only intrinsic noise generated from small fluctuations in concentration levels, small numbers of reactant molecules, and fast and slow reactions. Extrinsic noise is another source of stochasticity in gene regulation~\cite{Swain-Elowitz,St-Pierre}, and it would be interesting to see if this framework or a similar setup can be adapted to account for extrinsic stochasticity under the discrete approach. This framework also lends itself to the study of intrinsic noise and it is useful for the study of developmental robustness. For instance, one could ask
what the effect of this type of noise is on the dynamics of networks controlled by biologically inspired functions. 

Relating the propensity parameters to biologically meaningful information or having a systematic way for estimating them is very important. A preliminary analysis shows that it is possible to relate the propensity parameters in this framework with the propensity functions in the Gillespie algorithm under some conditions (see supporting material where for a simple degradation model, the degradation propensity is correlated by a linear equation with the decay rate of the species being degraded). More precisely, in the Gillespie algorithm~\cite{Gillespie,Gillespie2007}, if one discretizes the number of molecules of a chemical species into discrete expression levels such that within these levels the propensity functions for this species do not change significantly, then one obtains the setup of the framework presented here as a discrete model. That is, simulation within the framework presented here can be viewed as a further discretization of the Gillespie algorithm, in a setting that does not require exact knowledge of model parameters. For a similar approach see~\cite{Shunsuke}. 

\section*{Figures}
\newpage

\newpage

%

%
\end{document}